\documentclass[longnamesfirst,apjl]{emulateapj}
\usepackage{apjfonts,natbib,epsfig}
\tighten

\newcommand{\beq}{\begin{equation}}
\newcommand{\eeq}{\end{equation}}
\newcommand{\beqa}{\begin{eqnarray}}
\newcommand{\eeqa}{\end{eqnarray}}

\begin{document}

\title{
What is $L_{\star}$?: Anatomy of the Galaxy Luminosity Function
}

\author{Asantha Cooray$^1$ and Milo\v s Milosavljevi\'c$^2$}
\affil{$^1$Department of Physics and Astronomy, University of California, Irvine, CA 92617\\
$^2$Theoretical Astrophysics, California Institute of Technology, Pasadena, CA 91125}
\righthead{GALAXY LUMINOSITY FUNCTION}
\lefthead{COORAY \& MILOSAVLJEVI\'C}
\begin{abstract}

Using the empirical relations between the central galaxy luminosity and the halo mass, and between the 
total galaxy luminosity in a halo and the halo mass,
we construct the galaxy luminosity function (LF).
To the luminosity of the central galaxy in a halo of a given mass we assign log-normal scatter with 
a mean calibrated against the observations. In halos where the total galaxy luminosity exceeds
that of the central galaxy, satellite galaxies are distributed as a power-law in luminosity.
Combined with the halo mass function, this description reproduces the observed characteristics of
the galaxy LF, including a shape consistent with the Schechter function. 
When all galaxies are included, regardless of the environment or the Hubble type,
the Schechter $L_\star$ is the luminosity scale above which the central galaxy luminosity-halo mass relation flattens; $L_\star$ 
corresponds to $\sim 10^{13}M_{\sun}$ on the halo mass scale.  
In surveys where central galaxies in massive clusters are
neglected, either by design or because of the cosmic variance,
$L_\star$ is simply the mean luminosity of central galaxies in halos at the 
upper end of the selected mass range. The smooth, exponential decay of the 
Schechter function toward high luminosities reflects 
the intrinsic scatter in the central galaxy luminosity-halo mass relation. In addition to
the LF, the model successfully reproduces the
observed dependence of galaxy clustering bias on luminosity.


\end{abstract}

\section{Introduction}
\label{sec:introduction}

The mass function of dark matter halos is routinely measured in numerical simulations.
If one assumes that the galaxy luminosity is proportional the halo mass,
the abundance of galaxies at the faint and the bright ends of the luminosity 
range is significantly below the expected \citep{Vale:04,van:04}.
Semianalytic models of galaxy formation, including those
attempting to account for feedback and heating processes,
generally do not explain the shape of the galaxy luminosity function (LF; see, e.g., \citealt{Benson:03}).

In \citet{Cooray:05}, we studied the relation between 
the luminosity of the central galaxy and the mass of dark matter halo 
occupied by the galaxy (henceforth, the $L_{\rm c}$--$M$ relation).  We suggested that the
flattening of this relation in halos above $M_{\rm crit}\sim 10^{13}M_\odot$ 
is a consequence of the progressive decline in efficiency with 
which central galaxies accrete satellites in the course of hierarchical merging. 
We demonstrated that to explain the $L_{\rm c}$--$M$ relation, ongoing galaxy growth in massive halos 
via gas cooling and star formation need not be invoked.
Therefore, we depart from the classical
picture in which giant galaxies accrue mass continuously 
(e.g., \citealt{Rees:77,White:78}), and support the picture in which 
a bulk of the stellar mass is produced in smaller halos in which virial shocks do not raise the gas temperature to the virial temperature
\citep{Binney:04,Dekel:04}. 

Here, we demonstrate that the luminosity function of galaxies can be derived from two simple premises:
(1) The luminosities of central galaxies in halos of a given mass possess a log-normal intrinsic scatter 
with a mean and a dispersion identified with measured values, and (2) When the 
total luminosity of galaxies inside a halo exceeds that of the central galaxy, the luminosities of satellite galaxies are distributed as a power-law.
These two assumptions yield a LF of the 
\citet{Schechter:76} type, $\Phi(L)\propto L^{\alpha} e^{-L/L_\star}$, 
where $\alpha$ is the slope of the LF at the faint end, 
and $L_\star$ is usually thought of as a characteristic luminosity scale.  

Our construction of the LF is compatible in spirit with the ab initio synthesis of the 
conditional stellar mass function in \citet{Zheng:04}, and differs from the modeling of 
conditional luminosity function (CLF) in \citet{Yang:05}, where the Schechter form was assumed a priori.
Log-normal scatter of galaxy luminosities in {\it low-mass} halos 
was invoked by \citet{Yang:03b} to explain the scatter in the Tully-Fisher (TF) relation \citep{Tully:00}.
Here, we show that the log-normal scatter is crucial for explaining the
turnover of the Schechter LF at the {\it bright-end}.
Our approach is opposite to that of \citet{Vale:04}, who used the LF to extract information about the $L_{\rm c}$--$M$ relation.  We instead use the observationally determined $L_{\rm c}$--$M$ relation to reconstruct the LF.  We also calculate the CLF, and illustrate how the LF is built from the CLF.

In \S~\ref{sec:correlation}, we describe the construction of the galaxy LF.  In \S~\ref{sec:discussion},
we compare our LF with the observed LFs of cluster and field galaxies.
We adopt the 
current concordance cosmological model consistent with WMAP \citep{Spergel:03} with a Hubble constant of $h=0.7$ unless stated otherwise.

\section{Model Luminosity Function}
\label{sec:correlation}

\begin{figure*}[!t]
\centerline{\psfig{file=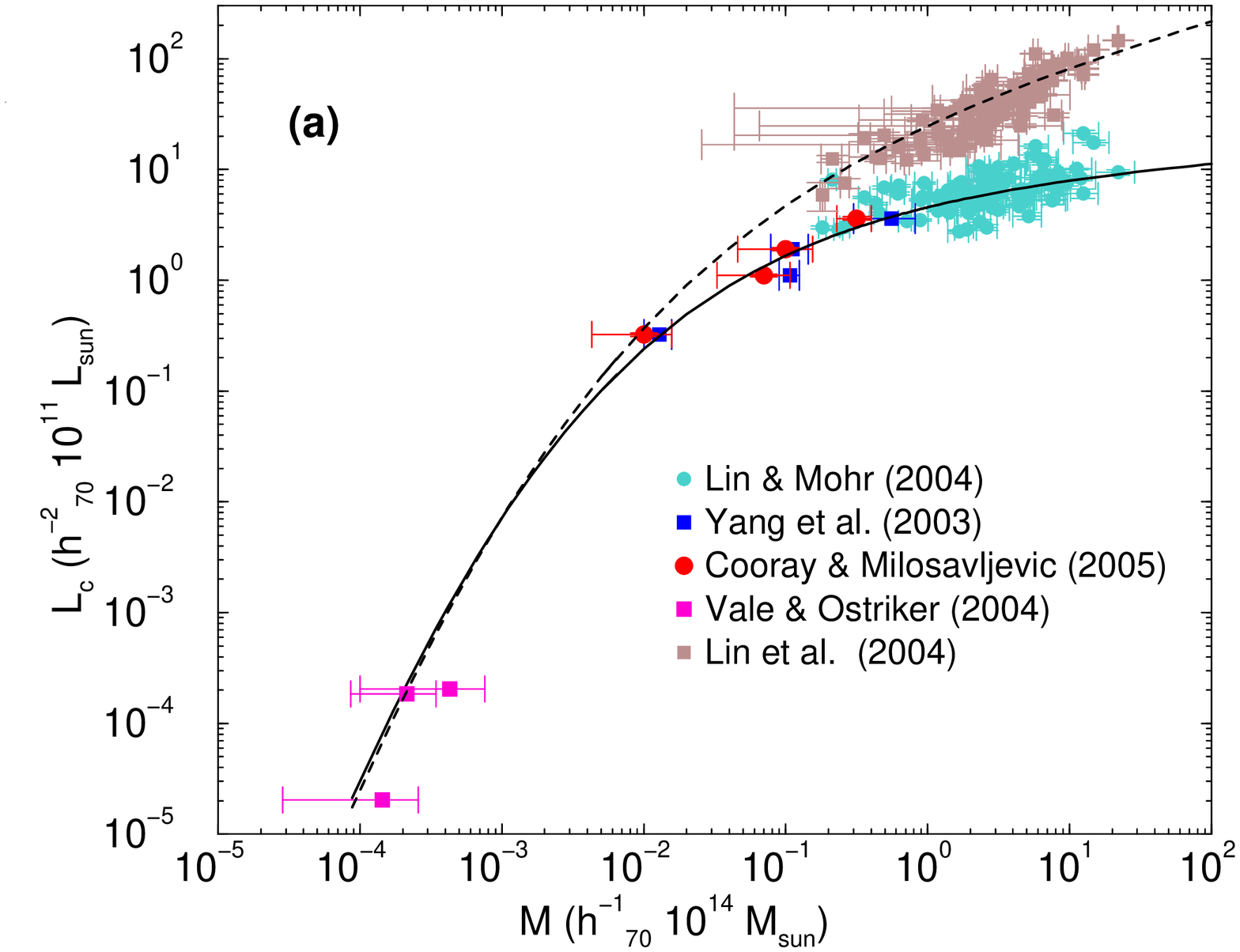,width=2.8in,angle=-90}
\psfig{file=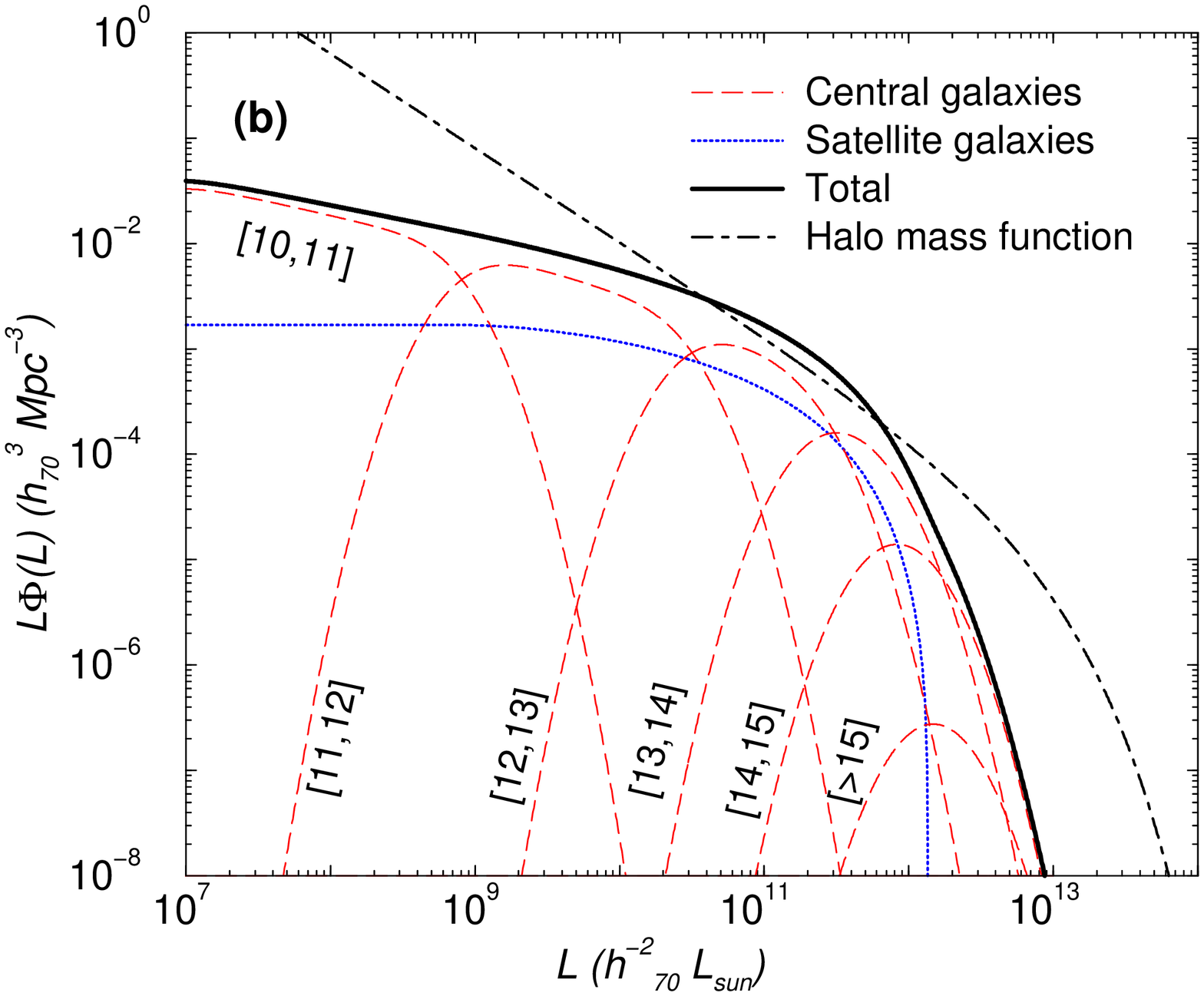,width=2.8in,angle=-90}}
\caption{(a) Central galaxy luminosity as a function of the
halo mass. The data are model fits to the SDSS galaxy-mass correlation function (\citealt{Cooray:05}; {\it red points}), 
the same masses estimated by Yang et al.\ (2003; {\it blue squares}), 
a direct measurement 
of galaxy luminosity and halo mass for a sample of galaxy groups and clusters from Lin \& Mohr (2004; {\it cyan points}).
 To extend the relation below the luminosities considered by \citet{Cooray:05}, we use the low luminosity data points from
Vale \& Ostriker (2004; {\it magenta squares}). We also show the total luminosity of galaxy groups and clusters based on
data of Lin et al.\ (2004; {\it small squares}).
(b) The LF of galaxies ({\it thick solid line}). We show the contributions to CLF from
central galaxies {\it dashed red lines}. From left to right, the CLFs
are shown for dark matter halos with masses separated into six decade intervals spanning
the range $(10^{10}-10^{16})M_\sun$ with
$\Sigma=0.25$. The total LF is shown as a solid black line.
The blue line is the 
total LF related to satellite galaxies in halo masses above $\sim 10^{11} M_{\sun}$
with $\gamma=-1$. Central
galaxies dominate the total LF at any luminosity.
For comparison, we also show the LF obtained assuming a constant mass-to-light
ratio $M/L=100$ ({\it dot-dashed line}). The total LF can be approximated by a Schechter function with $\alpha \approx -1.3$ 
and $L_\star\approx3\times10^{11}L_\sun$, where $\alpha$ is measured at $L=10^8L_\odot$.
The value of $L_\star$ shifts to lower luminosities when one 
ignores central galaxies in groups and clusters.}
\label{fig:sloan}
\end{figure*}

The CLF, denoted by $\Phi(L|M)$, is the average number of galaxies with luminosities between $L$ and $L+dL$ that reside in halos of mass $M$.  We separate the CLF into terms associated with central and satellite galaxies, $\Phi(L|M)=\Phi_{\rm c}(L|M)+\Phi_{\rm s}(L|M)$, where the central galaxy CLF is a log-normal distribution in luminosity with a mean $L_{\rm c}(M)$ and dispersion 
$\ln(10)\Sigma$
\begin{eqnarray}
\Phi_{\rm c}(L|M)  &=& \frac{ \Phi(M)}{\sqrt{2 \pi} \ln(10)\Sigma L} \exp \left\{-\frac{\log_{10} [L /L_{\rm c}(M)]^2}{2 \Sigma^2}\right\} 
\end{eqnarray}
and the satellite CLF is a power law $\Phi_{\rm s}(L|M) = A(M) L^{\gamma}$, where $\Phi(M)$ and $A(M)$ are normalization factors.  The separation into central and satellite galaxies is motivated by 
the halo model for galaxy statistics \citep{Cooray:02}.

The normalization $\Phi(M)$ of the central galaxy CLF is fixed such that $\int\Phi(L|M)LdL$ equals the average total luminosity of galaxies $L_{\rm tot}(M)$ in a halo of mass $M$. 
The normalization $A(M)$ of the satellite CLF can be obtained by defining $L_{\rm s}(M)\equiv L_{\rm tot}(M)-L_{\rm c}(M)$
and requiring that $L_{\rm s}(M)=\int_{L_{\rm min}}^{L_{\rm max}} \Phi_{\rm s}(L|M)LdL$, where the minimum luminosity of a satellite is $L_{\rm min}$,while we generally set $L_{\rm max}=L_{\rm c}$. 
We find below from an analysis the cluster LF that a more appropriate value for 
the maximal luminosity of satellites is between $L_{\rm c}/2$ and $L_{\rm c}/3$.
Our reconstruction is independent of 
the exact value assumed for $L_{\rm min}$, as long as it lies in the range $(10^6-10^8)L_{\sun}$. Note that 
$L_{\rm tot}(M)$ equals $L_{\rm c}(M)$ for $M<10^{11}M_{\sun}$, and thus $L_{\rm s}(M)=0$ on these scales.
The remaining two free parameters are
$\gamma$ and $\Sigma$; we discuss the determination of these parameters in \S~\ref{sec:discussion}.

We adopt the Sheth \& Tormen (1999; ST) mass function $dn/dM$ for dark matter halos as it approximates numerical simulations better than the \citet{Press:74} function (see, e.g., \citealt{Jenkins:01}). 
The usual, unconditional LF is then given by
\begin{equation}
\label{eq:lf}
\Phi(L) = \int_0^\infty \Phi(L|M) \frac{dn}{dM} dM .
\end{equation}
The CLF represents galaxy statistics better than the LF when wide-field data sets are available in which 
redshifts are 
measured for tens of thousands of galaxies
\citep{Yang:05}. 

For $L_{\rm c}(M)$, we make use of
the observed central galaxy luminosity measurements over six orders of magnitude in luminosity (see Fig.~\ref{fig:sloan}a).
Following \citet{Vale:04}, we employ a fitting function of the form
\begin{equation}
\label{eq:fitting}
L(M) = L_0 \frac{(M/M_1)^{a}}{[b+(M/M_1)^{cd}]^{1/d}}\, .
\end{equation}
For central luminosities, the parameters are $L_0=4.4\times10^{11} L_{\sun}$, $M_1=10^{11} M_{\sun}$,
$a=4.0$, $b=0.9$, $c=3.85$, and $d=0.1$. These values are different than in \citet{Vale:04}; their fit was constructed from the $b_J$-band luminosities of 2dF, while we use the $K$-band luminosities. For 
$L\gtrsim 5\times10^{10}L_{\sun}$, the relation is compatible with that derived in \citet{Cooray:05}.
For total luminosities, we also use the fitting formula in equation (\ref{eq:fitting}), but with 
$c=3.65$.  At the massive end, the total luminosity can alternatively be 
described as a power-law. The constructed LF does not change if power-law behavior is 
enforced there.
This is because the total LF (including field, group, and cluster galaxies) 
is dominated by central galaxies on any scale.
The overall shape of the LF is 
thus sensitive to the shape of the 
$L_{\rm c}$--$M$ relation.

\begin{figure*}[!t]
\centerline{\psfig{file=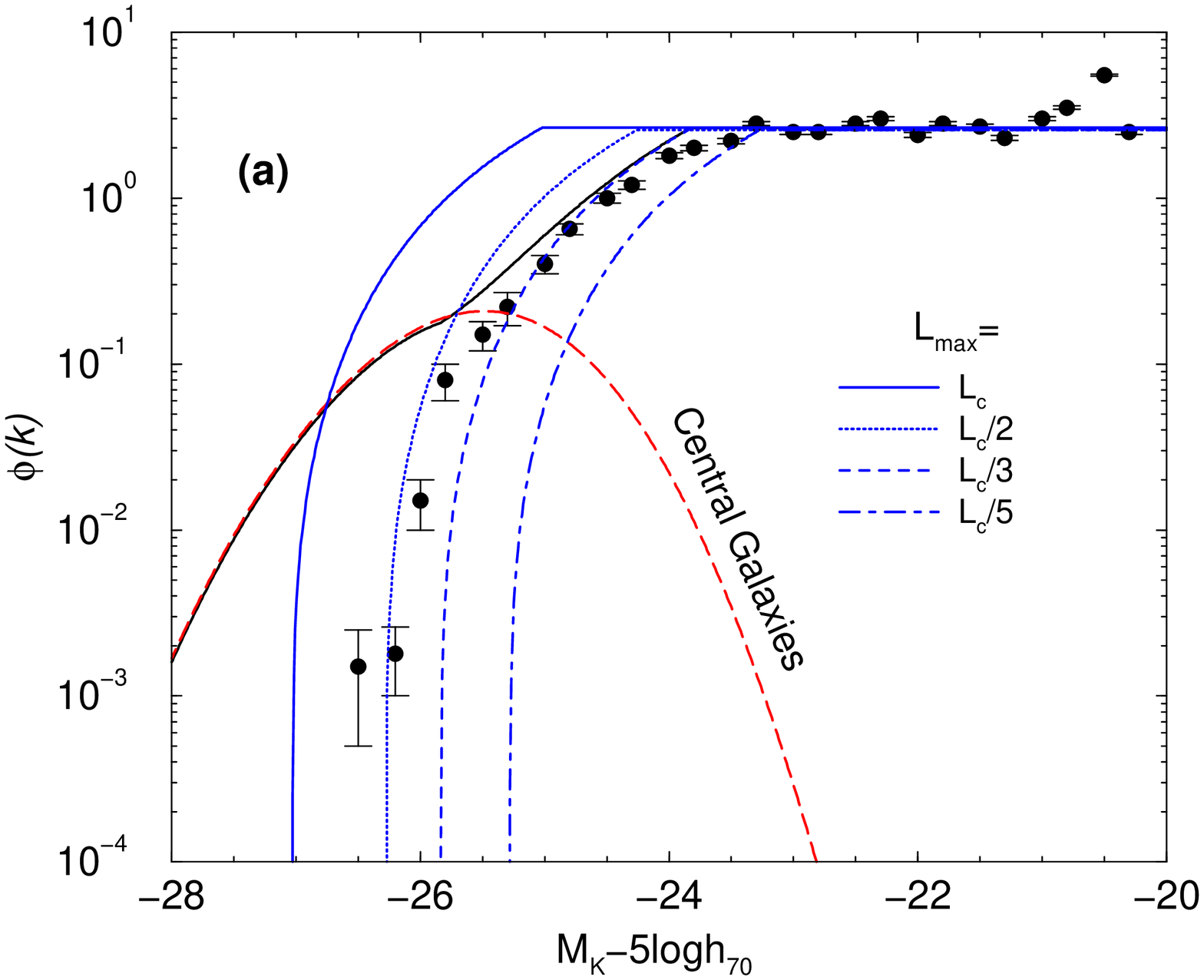,width=2.8in,angle=-90}
\psfig{file=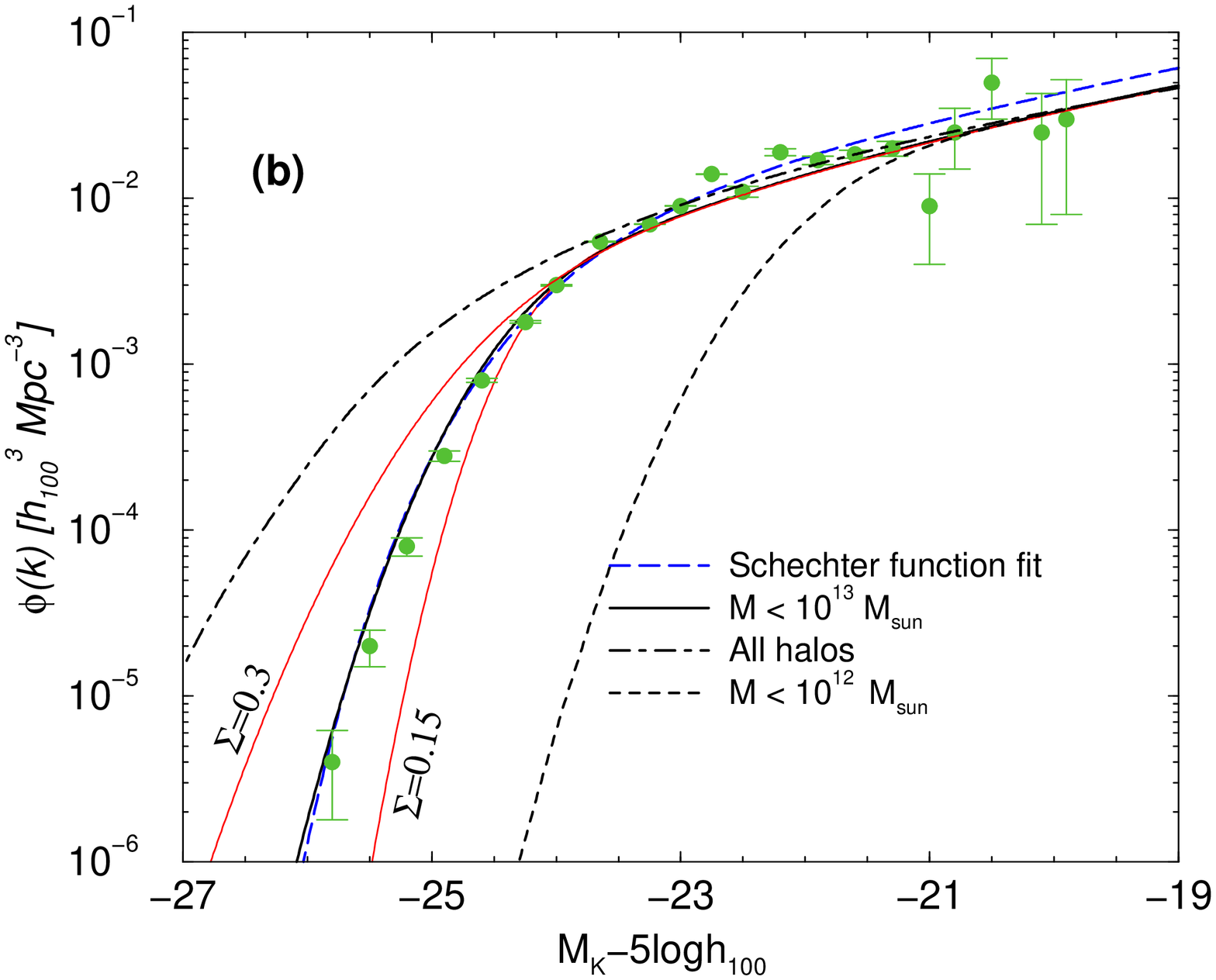,width=2.8in,angle=-90}}
\caption{(a) Galaxy cluster LF for $\gamma=-1$, where luminous central galaxies
are ignored in the construction ({\it blue lines}), for varying maximum luminosity of a satellite $L_{\rm max}=(1,1/2,1/3,1/5)L_{\rm c}$ ({\it blue lines}). The curves are arbitrarily normalized at the faint end of luminosity as the absolute normalization of $\phi(k)$ is unknown.
The data are from \citet{Lin:04}.  For reference, we also show the contribution to
the LF by central galaxies ({\it red dashed line}). The total LF clearly departs from the
Schechter form. 
 (b) The $K$-band field galaxy LF ({\it black lines}); we show the galaxy LF with maximum halo mass of $(0.1,1) \times 10^{13}M_\sun$ and scatter $\Sigma=0.23$. The dot-dashed line is the LF with all halos included.
The thin red lines with show the variation of the LF when
the width of the log-normal distribution, $\Sigma$, varies to $0.15$ and $0.3$, 
assuming a maximum mass of $10^{13} M_{\sun}$.  Note that $\Sigma\approx0.23$ describes the data best.  
The data are from \citet{Huang:03}. We also shows the Schechter function fit to the data in
\citet{Huang:03} with $M_{K\star}=-23.7$, $\alpha=-1.38$, and $\phi_\star=0.013 \textrm{ Mpc}^{-3}$ ({\it dashed blue line}).  Note that the Hubble constant $h=1$ in panel (b).
}
\label{fig:lum}
\end{figure*}

\section{Results and Discussion}
\label{sec:discussion}

In Figure \ref{fig:sloan}b, we show the constructed LF in which the galaxies
in halos in the mass range $(10^{10}-10^{16})M_\odot$ have been taken into account. 
We also plot the contributions to CLF from central galaxies 
and the contribution to LF from satellite galaxies.
The central galaxies dominate the LF on any 
luminosity scale, similar to the conclusion 
by \citet{Zheng:04} that central galaxies dominate the stellar mass function on any mass scale.
The fractional contribution of satellites to LF depends on the luminosity and has a maximum 
slightly below $L_\star \sim 3 \times 10^{11} L_\sun$.  
We now explore the origin of the slope $\alpha$ of the LF at the faint end, and then discuss the meaning of the characteristic scale $L_\star$ above which LF decreases steeply with increasing luminosity.

The faint-end slope of the LF is related to the 
low-mass slopes of the halo mass function and the $L_{\rm c}$--$M$ relation. 
At low halo masses, the mass function 
$dn/dM \propto M^{-2} \sigma(M)^{-\beta}$, where $\beta=1$ in the Press-Schechter mass function
and $\beta \approx0.4$ in the ST mass function. Here, $\sigma(M) \propto M^{-(n+3)/6}$ is the density variance on scales $M$, while $n$ is the slope of the matter power spectrum, $P(k) \propto k^n$.
Since central galaxies dominate the LF, we have $\Phi(L) \sim \int\Phi_{\rm c}(L|M) (dn/dM)dM$. 
If $L_{\rm } \propto M^\eta$, we can write $\Phi(L) \sim \int  
L^{\prime -1 -1/\eta + \beta (n+3)/6\eta} \delta(L'-L) dL'$, 
where we have ignored the scatter in the $L_{\rm c}$--$M$ relation by setting $\Sigma\rightarrow0$ and thus
replacing the log-normal distribution representing the dispersion of
central galaxy luminosities with a $\delta$-function at $L_{\rm c}$;
the faint-end slope of the LF is insensitive to $\Sigma$.
The faint-end LF then scales as $\Phi(L) \propto L^{-1-1/\eta + \beta (n+3)/6\eta}$.

\begin{figure}[!t]
\psfig{file=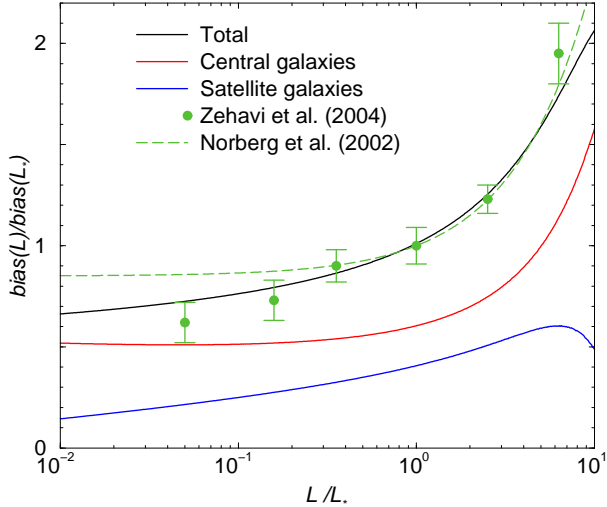,width=2.7in,angle=-90}
\caption{Galaxy bias as a function of luminosity calculated from conditional
luminosity functions ({\it black line}); we include the separate contributions from central galaxies ({\it red line}) and satellites ({\it blue line}). Also shown are the SDSS of \citet{Zehavi:04}, 
and a fit to the galaxy bias in 2dF data of \citet{Norberg:02} ({\it green dashed line}). 
}
\label{fig:bias}
\end{figure}

An examination of Figure \ref{fig:sloan}b shows that the faint-end slope of the LF is
defined by galaxies in dark matter halos with masses below $\sim
10^{11}M_\sun$. 
At these mass scales, $2 \lesssim \eta < 4$, where the upper limit corresponds to asymptotically low mass halos.
Since $-2\leq n\leq -1$, 
we have $\beta (n+3)/6\eta\ll 1$, and thus
$\Phi(L) \sim L^\alpha$ with $\alpha\gtrsim-1.5$. We expect $\alpha<-1.25$, unless the $L_{\rm c}$--$M$ relation is steeper than $\eta=4$ at the low-mass end.  
We expect that the slope of $\Phi(L)$ at the faint end is independent of the color selection and 
bandpass, as these choices only affect the normalization of the mass-to-light scaling.
In the case of the $K$-band galaxy LF, the measured faint-end slopes range from $\alpha=-1.39$ \citep{Huang:03}
to $\alpha=-0.93 \pm 0.04$ \citep{Cole:01}. 
At other wavelengths, the best-measured galaxy LF has $\alpha=-1.21 \pm 0.03$ in the $b_J$ band of
2dF \citep{Norberg:02} and the flatter
$\alpha=-1.05 \pm 0.01$ in the $r'$-band of SDSS \citep{Blanton:03}.  When the same luminosity function
was extended to fainter magnitudes, however, $\alpha=-1.3$ \citep{Blanton:04}, consistent with our expectations.

Provided that the $L_{\rm c}$--$M$ relation is well known on all scales, the LF of galaxies subject to a particular selection criterion based on galaxy type or environment (e.g., field galaxies, cluster galaxies, etc.) can be estimated directly by inserting the probability of selection in the integral in equation (\ref{eq:lf}).
Variation between selection criteria may be responsible for the observed variation in $\alpha$.  Selection criteria also seem to be responsible for the measurements with $\alpha \gtrsim -1$, which naively 
seem discrepant with the above predictions. If, e.g., a survey is biased toward ignoring the galaxies in low mass halos, one may indeed measure $\alpha\gtrsim -1$.  

It is unclear whether the faint-end slope of the LF changes with redshift.
\citet{Drory:03}, who studied the redshift evolution of the LF in the $K$-band, 
assumed a priori that $\alpha$ is independent of redshift.
Since the slope of the mass function at low masses 
is independent of redshift, any  
variation of $\alpha$ with redshift would imply that the $L_{\rm c}$--$M$ relation must also vary with redshift. 
The expected correlation between $\alpha$ and the slope of the $L_{\rm c}$--$M$ relation
can be tested directly if a galaxy survey, e.g., at $z\sim1$, can be combined with weak 
lensing mass estimates for galaxies in the same sample.

When an unbiased galaxy sample is used, $L_\star$ can be identified with the scale above which the luminosity scatter in the $L_{\rm c}$--$M$ relation dominates over the increase of luminosity with mass, i.e., $d\ln L_{\rm c}/d\ln M\sim \ln(10)\Sigma$. With $\Sigma \sim 0.25$, this yields 
$M_\star \sim 2 \times 10^{13}M_{\sun}$ and $L_\star=L_{\rm c}(M_\star)\sim 3 \times 10^{11} L_\odot$.  Above this value, the further increase in mean galaxy luminosity with halo mass is smaller than the scatter, 
and the corresponding CLFs overlap.
Had the $L_{\rm c}$--$M$ relation continued to increase at the rate $d\ln L_{\rm c}/d\ln M> \ln(10)\Sigma$,
the CLFs would have been disjoint and the Schechter function would not have exhibited the usual exponential 
cut-off.

Satellite galaxies do not contribute significantly to the overall 
LF. Nevertheless, galaxy cluster studies routinely measure the LF of satellite galaxies in clusters.
In Figure \ref{fig:lum}a, we plot the LF of satellite galaxies in clusters from \citet{Lin:04}.
Since the measurement is in the $K$-band and is based on the same data is in
Figure \ref{fig:sloan}a, a direct comparison with our model is possible. In Figure \ref{fig:lum}a, we plot the
LF with and without central galaxies, and also the LF of central galaxies alone.
In this calculation we assumed that the cluster mass lies in the range $(10^{13}-10^{15})M_\sun$ 
with a dispersion of $\Sigma=0.25$. Our construction reproduces the observational fact 
\citep{Lin:04} that the cluster LF, inclusive the central galaxies,
departs significantly from the Schechter form. Similarly, \citet{Trentham:02} 
decomposed the total LF of nearby clusters and groups into a log-normal and a Schechter function-like component. 
The \citet{Lin:04} data suggest that the power-law slope of the satellite distribution is $\gamma\approx -1\pm0.2$, while
the maximum
luminosity $L_{\rm max}$ of the satellite distribution is 
in the range $(1/3-1/2)L_{\rm c}$.

In Figure \ref{fig:lum}b, we compare the field galaxy LF of \citet{Huang:03} with the predictions of our model.
The definition of {\it field} galaxies here includes
all galaxies regardless of the environment. Nevertheless, their selection of galaxies may still somehow have been biased 
or affected by the cosmic variance (the random fluctuation in the number 
of groups and clusters in a survey of limited volume).
The \citet{Huang:03} data are best
described with a maximum mass of $\sim 10^{13}M_\sun$. This corresponding 
$L_\star$ is factor of 2 below the value expected based on the construction described in \S~\ref{sec:correlation}.
The discrepancy could perhaps be ascribed to the cosmic variance in the number of clusters in the sample of  \citet{Huang:03}. We vary the width $\Sigma$ of the Gaussian that describes the dispersion in luminosity at fixed halo mass.
To match the exponential behavior of the LF, $\Sigma\sim0.23$ is appropriate, 
which is perfectly consistent with the factor of $2$ vertical scatter in 
the $L_{\rm c}$--$M$ relation. This scatter is somewhat higher than 
$\Sigma \sim 0.17$ estimated by \citet{Yang:03b} for the bright-end of the TF relation (mass scales of $10^{13}M_\odot$); their estimate of the 
scatter was based on the data of \citet{Tully:00}.

As an example of a practical application of our reconstructed CLF, in Figure \ref{fig:bias}
we compare our prediction of the luminosity-dependent galaxy bias with the measurements from SDSS
\citep{Zehavi:04} and 2dF \citep{Norberg:02}.  The bias equals
$b(L) = \Phi(L)^{-1}\int b(M) \Phi(L|M)(dn/dM)dM $, 
where $b(M)$ is the halo bias based on the ST mass function \citep{Sheth:01}.
The bias increases monotonically from below unity 
to above unity as $L/L_\star$ ranges from $0.01$ to $10$.
This behavior is reproduced in our model which shows that the average bias is also
dominated by central galaxies at a given luminosity.

Semi-analytic studies have, thus far, not succeeded in reproducing the observed LF \citep{Benson:03}.
Here, we have presented a simple model that exposes the basic elements that shape the LF. The faint-end slope of the LF has 
received attention recently because it differs from the slope of the halo mass function \citep{Benson:02}.  
If the scaling of the average galaxy luminosity with the halo mass at luminosities below $L_\star$ can be explained,
LF can be directly recovered as we demonstrate in \S~\ref{sec:correlation}.  
The shape of the LF
at $L\gtrsim L_\star$ is then governed by the dissipationless merging model for luminosity growth presented in \citet{Cooray:05}.
There, we argued that feedback processes must prevent continued star formation on all luminosity scales $\gtrsim L_\star$.  
We also note that our construction may also make it easier to
reliably measure cosmological parameters from the galaxy LF, as first attempted in \citet{Seljak:02}.

\acknowledgements
A.~C.\ thanks members of Cosmology groups at Caltech and U.~C.\ Irvine for useful discussions.
M.~M.\ was supported at Caltech by a postdoctoral fellowship from the Sherman Fairchild Foundation.  
We thank Frank van den Bosch and Andrew Benson for helpful correspondence.


\begin{thebibliography}{}

\bibitem[Benson et al.(2002)]{Benson:02} Benson, A.~J., Lacey, C.~G., Baugh, C.~M., Cole, S., \& Frenk, C.~S.\ 2002, \mnras, 333, 156

\bibitem[Benson et al.(2003)]{Benson:03}
Benson, A.~J., Bower, R.~G., Frenk, C.~S., Lacey, C.~G., Baugh, C.~M., \& Cole, S.\ 2003, \apj, 599, 38

\bibitem[Binney(2004)]{Binney:04}
Binney, J.\ 2004, \mnras, 347, 1093

\bibitem[Blanton et al.(2003)]{Blanton:03}
Blanton, M.~R. et al.\ 2003, \apj, 819

\bibitem[Blanton et al.(2004)]{Blanton:04}
Blanton, M.~R., Lupton, R.~H., Schlegel, D.~J., Strauss, M.~A., Brinkmann, J., Fukugita, M., \& Loveday, J.\ 2004, preprint (astro-ph/0410164)

\bibitem[Cole et al.(2001)]{Cole:01}
Cole, S. et al.\ 2001, \mnras, 326, 255

\bibitem[Cooray \& Sheth(2002)]{Cooray:02}
Cooray, A., \& Sheth, R.\ 2002, Phys.\ Rep., 372, 1 (astro-ph/0206508)

\bibitem[Cooray \& Milosavljevi\'c(2005)]{Cooray:05}
Cooray, A., \& Milosavljevi\'c, M.\ 2005, preprint (astro-ph/0503596)

\bibitem[Dekel \& Birnboim(2004)]{Dekel:04}
Dekel, A., \& Birnboim, Y.\ 2004, preprint (astro-ph/0412300)

\bibitem[Drory et al.(2003)]{Drory:03} Drory, N., Bender, R., Feulner, G., Hopp, U., Marston, C., Snigula, J., \& Hill, G.~J.\ 2003, \apj, 595, 698

\bibitem[Huang et al.(2003)]{Huang:03}
Huang, J.-S., Glazebrook, K., Cowie, L.~L, \& Tinney, C.\ 2003, \apj, 584, 203

\bibitem[Jenkins et al.(2001)]{Jenkins:01}
Jenkins, A., Frenk, C.~S., White, S.~D.~M., Colberg, J.~M., Cole, S., Evrard, A.~E., Couchman, H.~M.~P., \& Yoshida, N.\ 2001, \mnras, 321, 372



\bibitem[Lin \& Mohr(2004)]{LinMohr:04}
Lin, Y., \& Mohr, J.~J.\ 2004, \apj, 617, 879

\bibitem[Lin, Mohr, \& Stanford(2004)]{Lin:04}
Lin, Y., Mohr, J.~J., \& Stanford, A.\ 2004, \apj, 610, 745

\bibitem[Norberg et al.(2001)]{Norberg:01}
Norberg, P., et al.\ 2001, \mnras, 332, 827

\bibitem[Norberg et al.(2002)]{Norberg:02}
Norberg, P., et al.\ 2002, \mnras, 336, 907

\bibitem[Press \& Schechter(1974)]{Press:74}
  Press, W. H., \& Schechter, P.\ 1974, \apj, 187, 425

\bibitem[Rees \& Ostriker(1977)]{Rees:77}
  Rees, M.~J., \& Ostriker, J.~P.\ 1977, \mnras, 179, 451



\bibitem[Schechter(1976)]{Schechter:76}
  Schechter, P.\ 1976, \apj, 203, 297

\bibitem[Seljak(2002)]{Seljak:02}
Seljak, U.\ 2002, \mnras, 334, 797

\bibitem[Sheth, Mo, \& Tormen(2001)]{Sheth:01}
Sheth, R.~K., Mo, H.~J., \& Tormen, G.\ 2001, \mnras, 323, 1

\bibitem[Sheth \& Tormen(1999)]{Sheth:99}
 Sheth, R.~K., \& Tormen, G.\ 1999, \mnras, 308, 119

\bibitem[Spergel et al.(2003)]{Spergel:03} Spergel, D.~N., et al.\ 2003, \apjs, 148, 175



\bibitem[Trentham \& Tully(2002)]{Trentham:02} Trentham, N., \& Tully, R.~B.\ 2002, \mnras, 335, 712

\bibitem[Tully \& Pierce(2000)]{Tully:00} Tully, R.~B., \& Pierce, M.~J.\ 2000, \apj, 533, 177

\bibitem[Vale \& Ostriker(2004)]{Vale:04}
Vale, A., \& Ostriker, J.~P.\ 2004, \mnras, 353, 189

\bibitem[van den Bosch, Yang, \& Mo(2004)]{van:04}
van den Bosch, F.~C., Yang, X., \& Mo, H.~J.\ 2004, preprint (astro-ph/0412018)



\bibitem[White \& Rees(1978)]{White:78}
 White, S.~D.~M., \& Rees, M.~J.\ 1978, \mnras, 183, 341

\bibitem[Yang et al.(2003a)]{Yang:03}
Yang, X., Mo, H.~J., Kauffmann, G., \& Chu, Y.~Q.\ 2003, \mnras, 339, 387

\bibitem[Yang, Mo, \& van den Bosch(2003)]{Yang:03b}
Yang, X., Mo, H.~J., \& van den Bosch, F.~C.\ 2003, \mnras, 339, 1057

\bibitem[Yang et al.(2005)]{Yang:05}
Yang, X., Mo, H.~J., Jing, Y.~P., \& van den Bosch, F.~C.\ 2005, \mnras, 358, 217


\bibitem[Zehavi et al.(2004)]{Zehavi:04}
  Zehavi, I., et al.\ 2004, preprint (astro-ph/0408569)

\bibitem[Zheng et al.(2004)]{Zheng:04}
  Zheng, Z., et al.\ 2004, preprint (astro-ph/0408564)



\end{thebibliography}
\end{document}